\documentclass[10pt, aps, amsmath, twocolumn, superscriptaddress, showpacs, prb, floatfix]{revtex4-2}

\usepackage{graphicx}
\usepackage{bm}
\usepackage{inputenc}
\usepackage{natbib}
\usepackage{braket}
\usepackage{verbatim}

\usepackage{xcolor}

\begin{document}

\title{Nonlinear Orbital and Spin Edelstein Effect in Centrosymmetric Metals}

\author{Insu Baek$^\dagger$}
\affiliation{Department of Physics, Pohang University of Science and Technology, Pohang 37673, Korea}
\author{Seungyun Han$^\dagger$}
\affiliation{Department of Physics, Pohang University of Science and Technology, Pohang 37673, Korea}
\author{Suik Cheon}
\affiliation{Department of Physics, Pohang University of Science and Technology, Pohang 37673, Korea}
\author{Hyun-Woo Lee}
\email{hwl@postech.ac.kr}
\affiliation{Department of Physics, Pohang University of Science and Technology, Pohang 37673, Korea}

\begin{abstract}

{\bf Abstract}

Nonlinear spintronics combines nonlinear dynamics with spintronics, opening up new possibilities beyond linear responses. A recent theoretical work [Xiao et al., Phys. Rev. Lett. 130, 166302 (2023)] predicts the nonlinear generation of spin density [nonlinear spin Edelstein effect (NSEE)] in centrosymmetric metals based on symmetry analysis combined with first principle calculation. However, its microscopic mechanism is limited to a specific set of materials with local inversion symmetry breaking and is not applicable to general materials. This paper focuses on the fundamental role of orbital degrees of freedom for the nonlinear generation in centrosymmetric systems. Using a combination of tight-binding model and density functional theory calculations, we demonstrate that nonlinear orbital density can arise independently of spin-orbit coupling. In contrast, spin density follows through spin-orbit coupling. We further elucidate the microscopic mechanism responsible for this phenomenon, which involves the NSEE induced by electric-field-induced orbital Rashba texture. In addition, we also explore the potential applications of the nonlinear orbital and spin Edelstein effect for field-free switching of magnetization.
\end{abstract}

\maketitle

\noindent {\bf Introduction}

Nonlinear responses to external perturbations exhibit distinct symmetry properties compared to linear responses. As a result, novel types of nonlinear responses can emerge that are not allowed in the linear regime, such as shift current and unidirectional magnetoresistance \cite{Shift_1, Shift_2, UMR_1, UMR_2, Cheon_2022}. Unlike charge, spin possesses directional degrees of freedom, and spin-direction-related phenomena forbidden in the linear regime may arise in the nonlinear regime. Nonlinear spintronics, an emerging field that integrates nonlinear dynamics with spintronics, has gained attention in recent studies \cite{NSHE_1, NSHE_2, SHE_photon, USMR_1, USMR_2}.

The orbital angular momentum (OAM) and spin densities can be generated from circular light \cite{IFE, IFE_1, NLE} or static electric field \cite{NREE_int, NREE_spin}. Especially, Ref. \cite{NREE_spin} predicts that in the second-order response, an external electric field can generate spin density even in centrosymmetric bulk systems, which is forbidden in the linear regime. Such nonlinear spin generation may be called the nonlinear spin Edelstein effect (NSEE) (instead of the Rashba-Edelstein effect \cite{REE}) since it occurs even in centrosymmetric systems where the Rashba-type spin-momentum coupling is forbidden \cite{REE_ISB}. The findings discussed in Ref.~\cite{NREE_spin} offer fresh insights from a fundamental physics standpoint. 

However, Ref.~\cite{NREE_spin} has a limitation. The model Hamiltonian used to elucidate the origin of NSEE requires inversion symmetry breaking at least locally. Although the model has the global inversion symmetry (therefore, the described system has hidden Rashba coupling~\cite{NREE_spin}), the inversion symmetry is broken locally, and the NSEE of the model vanishes when the local inversion symmetry is restored. This model does not provide a comprehensive explanation for the generic existence of NSEE, as the local inversion symmetry breaking is crucial. To go beyond this limitation and understand the NSEE in truly centrosymmetric systems (even at the local level), one needs to go beyond the spin degree of freedom, constructing a non-trivial Hamiltonian solely based on spin degrees of freedom is not feasible \cite{rt_2}.

Recently, there has been a growing recognition that spin dynamics are strongly influenced by orbital dynamics, which leads to the emerging field of orbitronics \cite{rt_3, rt_4, SHE_1, SHE_2, OT_1, ORB, ORE_1, ORE_2, han2023theory}. Orbital dynamics play a crucial role in spin dynamics, particularly in time-reversal centrosymmetric systems where only the orbital exhibits non-trivial dynamics, and spin dynamics is induced by the orbital dynamics through spin-orbit coupling (SOC). Hence, to achieve a comprehensive understanding of spin dynamics in centrosymmetric systems, it is desired to grasp orbital dynamics \cite{rt_2}. 

In this paper, we expand Ref.~\cite{NREE_spin} to include the orbital degree of freedom and investigate the NSEE and the nonlinear orbital Edelstein effect (NOEE) in centrosymmetric systems. First, we examine a centrosymmetric model system not only globally but also locally and demonstrate that it exhibits the NOEE even without SOC. In contrast, the NSEE vanishes in the zero SOC limit and becomes finite only when the SOC is turned on. Our study indicates that the orbital degree of freedom is crucial for the NSEE in centrosymmetric systems; the nonlinear orbital Edelstein effect arises first, which in turn induces the NSEE through the SOC. This clarifies the NSEE's mechanism and is our work's first main result. Second, we demonstrate that the second-order orbital Edelstein effect can be understood intuitively; one order of an electric field generates a momentum-dependent orbital moment, thereby converting a centrosymmetric system into an effective noncentrosymmetric system with the orbital Rashba effect. The other order of an electric field then induces the orbital Edelstein effect in the effective orbital Rashba system. Third, we perform density functional theory (DFT) calculations to establish the presence of NSEE and NOEE in a wide range of materials. To demonstrate this, we show that orthorhombic materials exhibit non-negligible NOEE as the NSEE. Furthermore, we illustrate that globally and locally centrosymmetric materials lacking NSEE in bulk due to $C_{3z}$ and $C_{4z}$ symmetries can have NSEE when sliced in specific orientations, which break those rotational symmetries. For instance, materials widely used in spintronics, such as fcc Pt, Rh, Ir, and Pd, can have non-negligible NSEE and NOEE if they are sliced along the (110) direction. Lastly, we remark that the polarization direction of spin/orbital current from the conventional spin/orbital Hall effect generated by the linear electric field and the polarization direction of the spin/orbital density derived from the NSEE generally exhibit disparities. This raises the possibility of harnessing this effect for the field-free switching of the magnetization in ferromagnets \cite{Shift_1}.

\begin{figure}[t]
\includegraphics[scale=0.27]{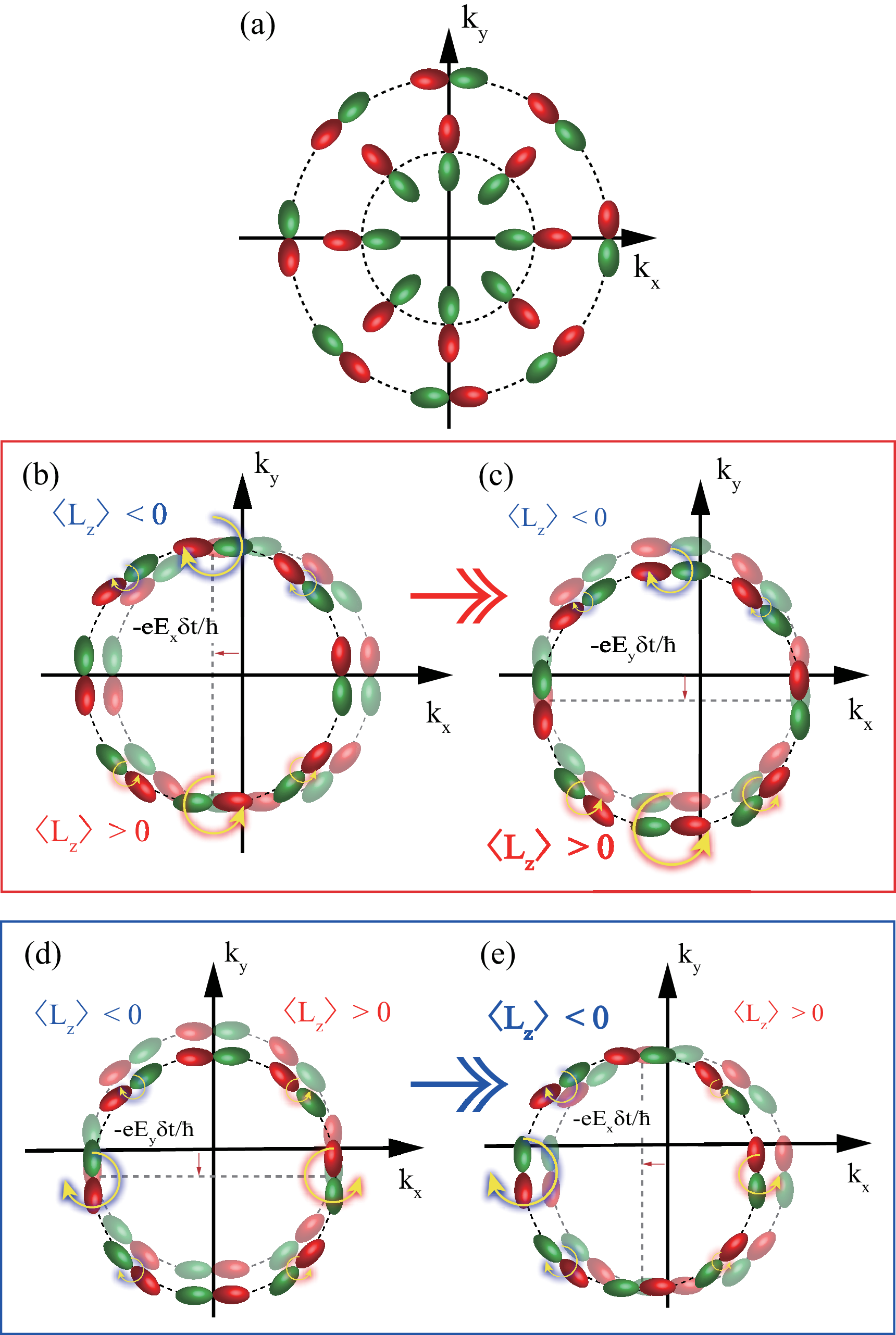}
 \caption{
 (a) The schematic diagram for real orbital texture. For other figures (b-e), the tangential (outer) band is only considered for a simple explanation. (b) The nonequilibrium orbital Rashba texture is initially induced by $E_x$. The transparent figure represents the equilibrium orbital texture. (c) Then, the nonequilibrium orbital Rashba texture is shifted by $E_y$ resulting the formation of nonzero OAM. The role of $E_x$ and $E_y$ can be exchanged so that the nonequilibrium orbital Rashba texture is generated by $E_y$ (d), and the NOEE is generated by $E_x$ (e). The sign of two contributions to the NOEE [(b) and (c) vs. (d) and (e)] are opposite, so they are canceled by each other when $C_{3z}$ or $C_{4z}$ symmetries exist. 
 }
\label{fig:1}
\end{figure}

\noindent {\bf Results}

{\bf Nonlinear orbital Edelstein effect.} We first consider the linear orbital Rashba-Edelstein effect (REE), which occurs in systems lacking inversion symmetry \cite{ORE_1, ORB, ORE_2, OT_2, ORE_3}. The eigenstates of such systems can have nonzero expectation values of the orbital angular momentum (OAM) that vary with \textbf{k}. Although the total OAM vanishes in equilibrium due to the time-reversal symmetry, which enforces $f(\varepsilon_{n\textbf{k}}) = f(\varepsilon_{\bar{n}-\textbf{k}})$ and $\bra{\psi_{n\textbf{k}}} \textbf{L} \ket{\psi_{n\textbf{k}}} = \bra{\psi_{\bar{n}-\textbf{k}}} \textbf{L} \ket{\psi_{\bar{n}-\textbf{k}}}$, where the state $(\bar{n}-\textbf{k})$ is the time-reversed state of the state $(n \textbf{k})$, the total OAM can acquire a finite value when an external electric field is applied since the field-induced Fermi surface shift is dissipative and breaks the time-reversal symmetry in the state occupation. Next, we consider our main systems where both inversion and time reversal symmetry coexist. In such systems, the OAM expectation values vanish for their eigenstates. Consequently, the field-induced Fermi surface shift alone cannot generate the net OAM density. The situation changes in the second-order response since the extra order of an electric field can induce interband mixing, thereby distorting the wavefunction. For the interband mixing, the orbital texture plays an important role \cite{rt_3}. To illustrate this point, we note that energy eigenstates with an expectation value of zero for OAM, such as $p_x$ and $p_y$ orbitals, vary across \textbf{k}-space \cite{rt_3, rt_1}. When an electric field is applied, each eigenstate gets deformed ($\ket{\psi_{n\textbf{k}}} \rightarrow \ket{\tilde{\psi}_{n\textbf{k}}}$) due to the real orbital texture \cite{rt_1, rt_3} and acquires \textbf{k}-dependent nonzero OAM expectation value [Fig.~\ref{fig:1}(b)]. Roughly speaking, the acquired OAM $\textbf{L}$ is proportional to $\textbf{E} \times \textbf{k}$ and thus generates the orbital Hall effect, where $\textbf{L}$, $\textbf{E}$, and $\textbf{v}$ are perpendicular to each other. If we write this field-induced effect into a perturbation equation up to the first order, we obtain

\begin{equation}\label{eqn:1}
    \mathcal{H} =\mathcal{H}_{eq}+ \alpha(\textbf{k}) (\textbf{k} \times \textbf{L}) \cdot \textbf{E} + \mathcal{O}(\textbf{E}^2),
\end{equation}
which is derived in \cite{go2023}. This can be interpreted as forming an electric field-induced orbital Rashba texture resulting from breaking inversion symmetry by the applied electric field [Figs.~\ref{fig:1} (b) and (d)]. For instance, when an electric field is applied to a two-dimensional system along the \textit{x} direction, it gives rise to an $L_z$ orbital current flowing along the \textit{y}-direction, which can be expressed as $\alpha(k_y L_z) E_x$. This phenomenon can be understood as forming an orbital Rashba texture induced by breaking inversion symmetry along the $x$ direction by the electric field. If an additional electric field is applied along the \textit{y}-direction to the system where the inversion symmetry is effectively broken by an electric field along the \textit{x}-direction [Eq. (\ref{eqn:1})], the \textit{y}-direction field effectively breaks the time-reversal symmetry through the Fermi surface shift and gives rise to a finite total OAM [Fig.~\ref{fig:1}(c)]. This intriguing observation corresponds to NOEE, which will be investigated in this paper. In the case of NSEE, it emerges through the conversion of NOEE by SOC. To achieve a giant NSEE, a crucial requirement is the presence of a substantial NOEE. Without such NOEE, achieving a giant NSEE becomes difficult. The reason lies in the role of SOC, which is crucial for generating NSEE. Considering that SOC is given as $\textbf{L} \cdot \textbf{S}$ in centrosymmetric systems, the effect of SOC is more pronounced when the $\textbf{L}$ is large (in the case of a substantial NOEE), and conversely if the $\textbf{L}$ is small, the influence of SOC diminishes. Therefore, to make the giant NSEE, giant NOEE is primary. 

Deriving this phenomenon using the Moyal-Keldysh \cite{Moyal} formula provides a formal expression that captures the physics described earlier, which is given by

\begin{subequations}\label{eqn:a9}
\begin{equation}\label{eqn:a9a}
     \delta X_z = \chi^{X_z}_{ij} E_i E_j = e^2 \hbar \tau \alpha^{X}_{zij} E_i E_j
\end{equation}
\begin{equation}\label{eqn:a9b}
    \chi^{X_z}_{ij} = e^2 \hbar \tau \sum\limits_{n \neq m} \int \frac{d^d \textbf{k}}{(2 \pi)^d} f(\varepsilon_{n\textbf{k}}) \Big( \frac{\partial \Omega_{nm\textbf{k}}^{i}}{\partial k_j} + \frac{\partial \Omega_{nm\textbf{k}}^{j}}{\partial k_i} \Big),
\end{equation}
\begin{equation}\label{eqn:a9c}
    \Omega_{nm\textbf{k}}^{i} = \operatorname{Im}\Big[ \frac{\bra{u_{n\textbf{k}}}\hat{X}_z\ket{u_{m\textbf{k}}}\bra{u_{m\textbf{k}}}{v_i}\ket{u_{n\textbf{k}}}}{(\varepsilon_{n\textbf{k}} - \varepsilon_{m\textbf{k}} - i \Gamma)^2}\Big],
\end{equation}
\end{subequations}
where $\Gamma$ is the energy level-broadening constant and $\textbf{X}$ is either $\textbf{L}$ or $\textbf{S}$. It allows us to quantitatively evaluate the interplay between the electric field-induced orbital Rashba texture [$\Omega_{nm\textbf{k}}^{i}$ (Fig.~\ref{fig:1}(b)) and $\Omega_{nm\textbf{k}}^{j}$ (Fig.~\ref{fig:1}(d))] and Fermi surface shift [$\frac{\partial f(\varepsilon_{n\textbf{k}})}{\partial k_j}$ (Fig.~\ref{fig:1}(c)) and $\frac{\partial f(\varepsilon_{n\textbf{k}})}{\partial k_i}$ (Fig.~\ref{fig:1}(e))] to generate the resulting nonzero OAM density. A few crucial remarks follow. Firstly, it is worth noting that the OAM density derived by this represents the only term in the time-reversal and inversion symmetric metals \cite{NREE_spin}. Next, for Eq.~(\ref{eqn:a9b}) to yield a nonzero result, both $C_{3z}$ and $C_{4z}$ symmetries must be broken since, otherwise, the two terms in Eq.~(\ref{eqn:a9b}) cancel out each other. Hence, in order to achieve a substantial NOEE, we have to focus on materials that significantly break these symmetries. Moreover, the nonzero OAM density resulting from this is roughly proportional to the anisotropy of $\alpha(\textbf{k})$ in \textbf{k}-space, which is linked to the anisotropy of the orbital Hall conductivity (OHC) in linear order induced by $E_i$ or $E_j$. That is, the degree of anisotropy of the OHC governs the magnitude of the NOEE. We note that this concept aligns with the notion of anomalous spin polarizability as introduced in Ref.~\cite{NREE_spin}.

\begin{figure}[t]
\includegraphics[scale=0.45]{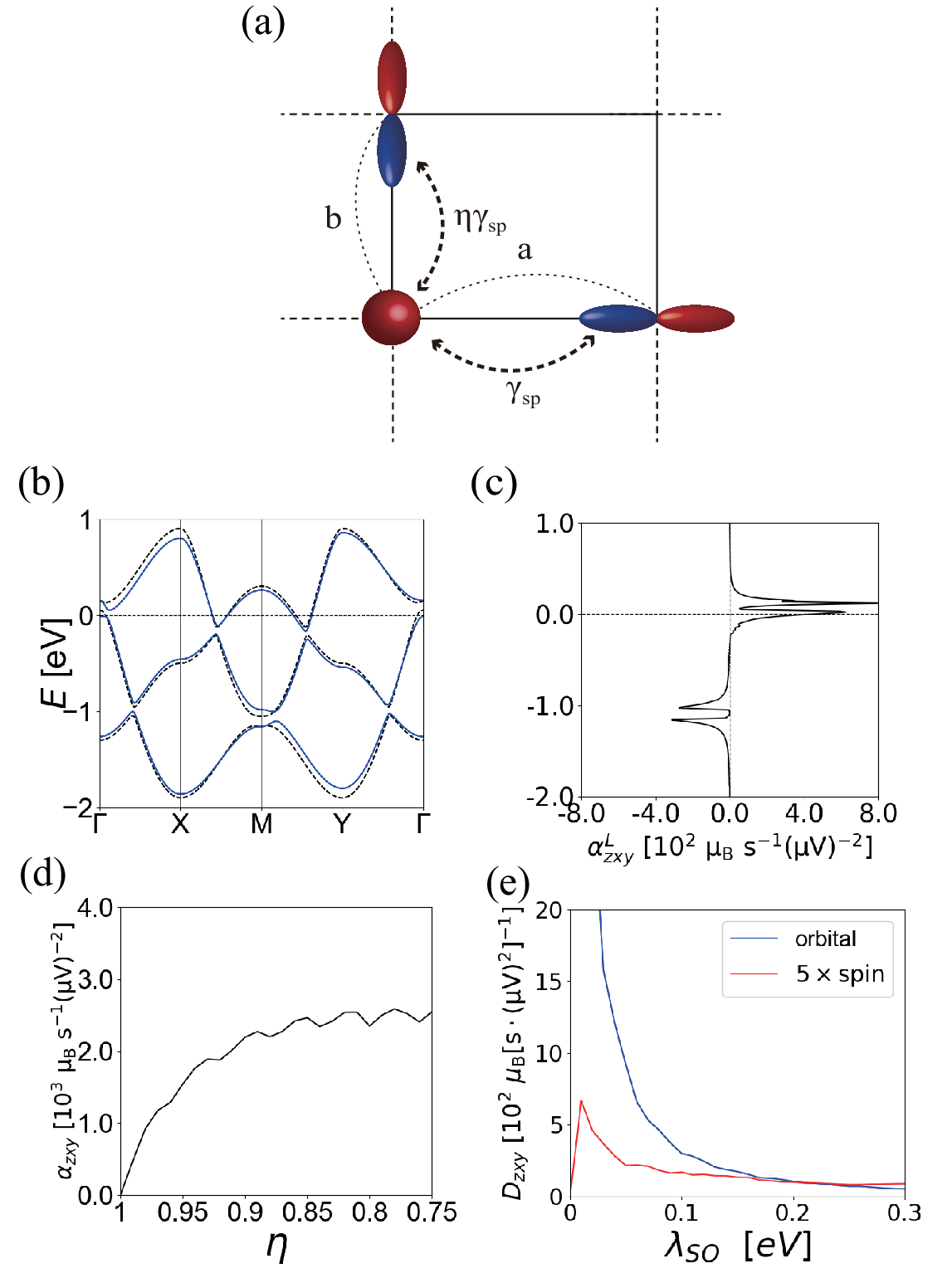}
\caption{
(a) Hopping rules of rectangular lattice with $s$, $p_x$, and $p_y$ orbitals model ($p_z$ orbitals are omitted). Since the lattice constants are different, the strength of $sp$ hybridization is different, which is captured by the parameter $\eta$. (b) The band structure of Eq.~(\ref{tightbinding}) with $\eta=1$ (dotted lines) and $\eta = 0.9$ (blue lines). Three bands correspond to three $p$-orbital character dominant bands, while the $s$-orbital character dominant band is omitted since it's far away from $p$-orbital character bands. (c) Result in NOEE from tight-binding calculation with $\eta = 0.9$. Near the exact crossing point (around 0 eV and $-1$ eV), the OAM density shows significant growth. (d) The nonlinear OAM accumulation $\alpha^{L}_{zxy}$ with respect to the anisotropy $\eta$. (e) The nonlinear OAM $\alpha^{L}_{zxy}$ and spin $\alpha^{S}_{zxy}$ accumulation with respect to the SOC $\lambda_{\rm{SO}}$.
}
\label{fig:2}
\end{figure}

{\bf Tight-binding model illustration.} To illustrate the physics described above, we introduce an intuitive tight-binding model. Our analysis considers a rectangular lattice with $s$, $p_x$, and $p_y$ orbitals with nearest-neighbor hopping [Fig.~\ref{fig:2}(b)]. The $C_{3z}$ rotation symmetry is obviously absent, and the $C_{4z}$ rotation symmetry is broken in the lattice spacing \textit{a} and \textit{b} along the \textit{x} and \textit{y} directions are different from each other. The hopping rule is schematically depicted in Fig.~\ref{fig:2}(a). Here, $\eta$ is the parameter for $C_{4z}$ rotational symmetry breaking. Note, however, that the model system has the inversion symmetry both locally and globally.

 Figure \ref{fig:2} illustrates the band structure calculated at $\eta=1$ and $\eta = 0.9$ and results obtained by calculating Eq.~(\ref{eqn:a9}) using the tight-binding model. In the case of a square lattice ($a=b$, $\eta=1$), where $C_{4z}$ symmetry is preserved, the NOEE caused by the three types of the quadratic perturbations to $E_x^2$, $E_y^2$ and $E_xE_y$ fields are all zero. However, as the value of $\eta$ deviates from 1, NOEE response to $E_xE_y$ yields nonzero values, and the overall OAM density increases [Fig.~\ref{fig:2}(c)]. This is due to the increasing anisotropy of the \textbf{k}-dependent orbital angular momentum derived from the linear order as $\eta$ moves further away from 1. Figure~\ref{fig:2}(d) shows that the OAM density increases as the $C_{4z}$ symmetry is broken. The NOEE due to $E_x^2$ and $E_y^2$ perturbation is still zero for this case due to the mirror $M_x$ and $M_y$ symmetry, respectively. From the comparison with the band structure [Fig.~\ref{fig:2}(b)], it becomes evident that the NOEE arises mainly near the band touching points. The nonlinear spin density is proportional to the SOC strength $\lambda_{\rm{SO}}$ [Fig.~\ref{fig:2}(e)]. Both the OAM and spin densities become smaller as $\lambda_{\rm{SO}}$ increases since the band gap, which generates the OAM and spin densities increases as $\lambda_{\rm{SO}}$ increases. An intriguing observation is that the NOEE and NSEE emerge even in this simplified model featuring distinct hopping parameters along the \textit{x}-axis and \textit{y}-axis. This suggests that the phenomenon may be prevalent in a wide range of multi-orbital systems exhibiting similar symmetry-breaking characteristics, providing insight into materials that can have large NOEE and NSEE. 

{\bf DFT calculation.} In this section, we calculate both NOEE and NSEE in real materials with the aforementioned symmetries broken. In centrosymmetric orthorhombic metals, such as IrW, MoRh, MoIr, and TiPt \cite{Ref_IrW, Ref_TiPt}, we find that when the electric field $E = 10^5 \ \rm{V} / \rm{m}$ \cite{Efield_1, Efield_2} is applied, the NOEE-induced OAM and NSEE-induced spin density are given by $\delta L_z$ and $\delta S_z \sim 10^{-7} - 10^{-8} \ \mu_{\rm{B}} / (\rm{nm})^3$. We also find that even the materials that do not exhibit the NOEE and the NSEE in bulk can have the NOEE and the NSEE when they are sliced into films in the proper direction. Materials widely used in spintronics, such as fcc Pt, Rh, Pd, Ir, and bcc V \cite{SHE_PtTa, Ref_SHA_Ir, SHE_Rh, SOT_Pd, SHE_V}, can have non-negligible NOEE and NSEE if it is sliced along the (110) direction since they only have $C_{2z}$ rotational symmetry. Leaving the calculation details in the supplementary, we evaluate those materials' NOEE-induced OAM and NSEE-induced spin densities and summarize them into Table \Ref {table:1}. When the electric field $E = 10^5 \ \rm{V} / \rm{m}$ \cite{Efield_1, Efield_2} is applied, the NOEE-induced OAM and NSEE-induced spin density are given, $\delta L_z$ and $\delta S_z \sim 10^{-7} - 10^{-8} \ \mu_{\rm{B}} / (\rm{nm})^3$ at the true Fermi energy in room temperature with thickness $\approx 3 \ \rm{nm}$. We also find that with doping, NSEE values of Pd and Ir can be enlarged \cite{supple}.

\begin{table}[t]
\caption{
The NOEE-induced OAM density $\delta L_z$ and spin density $\delta S_z$ with unit $10^{-7} \ \mu_{\rm{B}} / (\rm{nm})^3$ at the true Fermi energy for 11-layered films from DFT calculation at T = 300 \rm{K}. The external electric field is set as $E = 10^5 \ \rm{V} / \rm{m}$ \cite{Efield_1, Efield_2}. The relaxation time is $ \tau = 10 \ \rm{fs}$.
}
\begin{ruledtabular}
\begin{tabular}{l c c}
            &    $\delta L_z \ [10^{-7} \mu_{\rm{B}} / (\rm{nm})^3]$   &    $\delta S_z \ [10^{-7} \mu_{\rm{B}} / (\rm{nm})^3]$   \\
\hline
Bulk    &   &   \\
\hline
    IrW   &       $0.641$       &        $0.403$      \\
    MoIr  &       $0.316$       &        $0.851$      \\
    MoRh  &      $-0.417$       &        $0.128$      \\
    TiPt  &      $-0.107$       &       $-2.534$      \\
\hline
(110) films &       &                   \\
\hline
    Pt      &      $-1.466$       &       $-3.531$      \\
    Pd      &       $1.162$       &       $-2.822$      \\
    Ir      &      $-1.770$       &        $3.014$      \\
    Rh      &      $-3.796$       &       $-0.410$      \\
\end{tabular}
\label{table:1}
\end{ruledtabular}
\end{table}

\noindent {\bf Discussion } 

\begin{figure}[t]
 \includegraphics[scale=0.27]{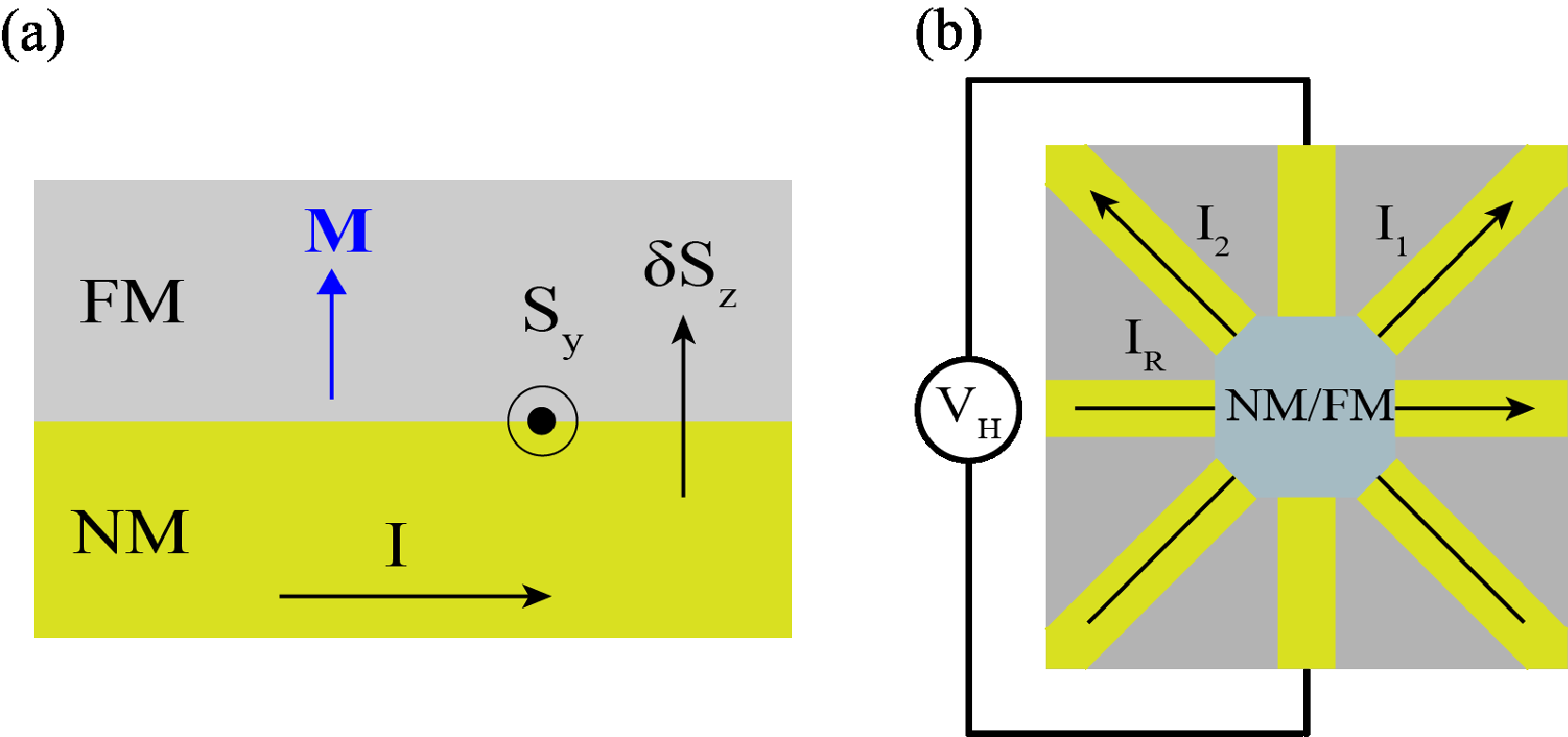}
 \caption{
 (a) The microscopic procedure for the field-free switching in a nonmagnet/ferromagnet bilayer.
 (b) The schematic diagram for the field-free switching measurement.
 }
 \label{fig:4}
 \end{figure}

In this section, we explore potential applications of the NSEE, particularly its use in field-free switching. We propose the feasibility of field-free switching utilizing the NSEE generated by centrosymmetric metals. In a typical experimental setup for switching, an electric field is applied to the FM/heavy metal (HM) structure, where the spin current from the spin Hall effect in the HM exerts an influence on the FM's magnetization [Fig.~\ref{fig:4}(a)]. Traditionally, deterministic switching requires an external magnetic field \cite{Field_Free_4, Field_Free_5, Field_Free_6, Field_Free_7} or mirror symmetry breaking of the sample \cite{Field_Free_1, Field_Free_8, Field_Free_9, Field_Free_10, Field_Free_3}. However, with the materials we have proposed, an alternative scheme to realize the field-free deterministic switching is possible since the second-order effect of the electric field generates spin density with a polarization distinct from that originating from the spin Hall effect. This spin density can diffuse into FM and can serve a similar function, potentially replacing the need for an external magnetic field [Fig.~\ref{fig:4}(a)]. To validate the scenario described above, the spin density generated in the NM material must be transferred to the FM. To investigate this, we truncate the NM material to a finite size to observe if the NSEE was induced at the sample's edge. The results \cite{supple} showed that the NSEE phenomenon occurred not only within the bulk of the material but also at the sample's edge with the comparable size of the bulk value. This indirectly suggests that the spin density generated at the NM sample's edge due to the NSEE in the NM/FM structure can effectively diffuse into the FM material. This implies that the NSEE-induced spin/NOEE-induced OAM density at the edges plays an equally important role at the edges as the spin/OAM density induced by conventional linear REE. 

Now, we provide a more detailed explanation of an experimental setting for field-free switching. Figure \ref{fig:4}(b) illustrates the experimental setup designed to capture the characteristics mentioned above for the field-free switching effect. We consider a situation where the magnetism is switched in the opposite direction when it is out-of-plane in equilibrium ($+z$ or $-z$ direction). First, the magnetization is reoriented to lie along in-plane direction by the spin/orbital torque generated by the spin/orbital Hall effect induced by the linear order of the electric field \cite{Field_Free_1}. Without the NSEE, the probability that the magnetization is returned to its initial direction or flopped is equal. However, with the NSEE, $\delta S_z$ is generated [Fig.~\ref{fig:4}(a)], which makes deterministic switching. Furthermore, using the property that the sign of the spin induced by NSEE depends on the current path [$I_1$ vs. $I_2$ in Fig.~\ref{fig:4}(b)], it becomes possible to achieve the field-free switching in both directions ($+z$ to $-z$ and $-z$ to $+z$). Since the NSEE and NOEE have $\sin(2\phi)$ dependence, the out-of-plane spin/orbital density induced by NSEE/NOEE is opposite for $I_1$ and $I_2$. Specifically, when an electric field is applied along the $I_1$ direction and induces an out-of-plane density of $+z$, the equilibrium magnetization can only be switched when it is initially along the $-z$ direction. On the other hand, if the magnetization is initially oriented in the $+z$ direction, it will return to its original direction. In the $I_2$ direction, the opposite is true. This current-path-dependent field-free switching property serves as a characteristic that can experimentally verify the presence of NSEE.

\noindent {\bf Conclusion}

In summary, we have identified the microscopic origin of the NSEE and dealt with the technical issue of the energy level broadening by which the NSEE is influenced severely. For the microscopic origin, we have elucidated that the orbital degree of freedom fundamentally underpins the origin of the NSEE. For the energy level broadening issue, we have applied the energy level broadening to the calculations and found that the NSEE is sizable only at low temperatures. The substantial NSEE that we have discovered is anticipated to give rise to significant dynamics and pave the way for diverse phenomena like field-free switching. This research stands as a bridge between nonlinear dynamics and nonlinear spin/orbitronics, opening up exciting prospects for both fundamental understanding and practical applications.

\noindent {\bf Methods}

{\bf Details of the Tight-binding Hamiltonian.} The tight-binding Hamiltonian in the ($s$, $p_x$, $p_y$, $p_z$) basis is given by

\begin{align}\label{tightbinding}
    \mathcal{H} = 
     \begin{bmatrix}
        H_{ss} & H_{sp_x} & H_{sp_y} & 0 \\
        H_{sp_x}^{*} & H_{p_xp_x} & 0 & 0 \\
        H_{sp_y}^{*} & 0 & H_{p_yp_y} & 0 \\
        0 & 0 & 0 & H_{p_zp_z}
    \end{bmatrix},
\end{align}
where $H_{ss}=E_s + 2t_s[\cos(k_x a)+\eta \cos(k_y a)]$, $H_{sp_x} = 2i\gamma_{sp}\sin(ak_x)$, $H_{sp_y} = 2i\eta\gamma_{sp}\sin(bk_y)$, $H_{p_xp_x}=E_p+2t_{pp\sigma}\cos(k_x a)+2\eta t_{pp\pi}\cos(k_y b)$, $H_{p_yp_y}=E_p+2t_{pp\pi}\cos(k_x a)+2\eta t_{pp\sigma}\cos(k_y b)$, $H_{p_zp_z}=E_p+2t_{pp\pi}[\cos(k_x a)+2\eta \cos(k_y b)]$. We choose $E_s = 3.0$, $E_p = t_s = \gamma_{sp} = t_{pp \sigma} = -0.5$, $t_{pp \pi} = 0.2$ in eV. Note that the magnitude of the $sp$ hybridization ($\gamma_{sp}$), inter-\textit{p}-orbital hopping parameters ($t_{pp\sigma}$, and $t_{pp\pi}$) varies between \textit{x} and \textit{y} directions by the parameter $\eta$. Thus, the difference $\eta - 1$ may be regarded as the strength of the $C_{4z}$ rotation symmetry breaking. We consider two $\eta$ values ($\eta=$ 1 and 0.9).

{\bf Details of the First Principle Calculation.} We proceed with the full-potential DFT calculation as follows. First, the self-consistent electronic structures are obtained using the full-potential linearization augmented wave method \cite{FLAPW} with code FLEUR \cite{FLEUR}. We use Perdew-Burke-Ernzerhof exchange-correlation functional within the generalized gradient approximation \cite{PBE}. The Brillouin zone is sampled using the $16 \times 16 \times 16$ Monkhorst-Pack \textbf{k}-point mesh for bulks and the $32 \times 32 \times 1$ Monkhorst-Pack \textbf{k}-point mesh for films, respectively \cite{MPgrid}. The muffin-tin radii of each atom and the plane wave cutoffs were set to $2.3 \ a_0$ and $5.0 \ a_0^{-1}$ (where $a_0$ is Bohr radius) for all structures. 

Next, we obtain the maximally localized Wannier functions (MLWFs) from the Bloch states with WANNIER90 code \cite{WANNIER90}. The Brillouin zones are sampled with the equidistant $8 \times 8 \times 8$ \textbf{k}-mesh for bulks and with the equidistant $8 \times 8 \times 1$ \textbf{k}-mesh for films respectively, including the $\Gamma$ point. Bloch states are initially projected into $s, \ p_x, \ p_y, \ p_z, \ d_{xy}, \ d_{yz}, \ d_{xz}, \ d_{x^2 - y^2}$, and $d_{z^2}$ states. We obtained 18 MLWFs out of 36 bands for each atomic site. The frozen windows are set to include a region 5 eV higher than the true Fermi energy. 
We employ the Kubo Formula within quadratic response theory to evaluate the induced orbital and spin magnetization derived in the previous section. The \textbf{k} integration in Eq. (\ref{eqn:a9}) is calculated using uniformly distributed 240 $\times$ 240 $\times$ 240 \textbf{k}-mesh grid for bulks and uniformly distributed 480 $\times$ 480 \textbf{k}-mesh grid for films, respectively. The temperature is set to $T = 300 \ \rm{K}$. The nonlinear OAM and spin accumulation are calculated by increasing the Fermi energy $\varepsilon_{\rm{F}}$ from $-2 \ \rm{eV}$ to $2 \ \rm{eV}$ with $0.01 \ \rm{eV}$ step, where the true Fermi energy is set to $\varepsilon_{\rm{F}} = 0$.

\noindent {\bf Data Availability}

Data in the manuscript can be provided by the corresponding author on request.

\noindent {\bf Author Contributions}

H.-W. Lee and S. Han conceived this work. I. Baek and S. Han contributed equally to this work. I. Baek derived the formula, performed the first principle calculations, and wrote the result part of the manuscript under the supervision of H.-W. Lee. S. Han performed the tight-binding model calculation and wrote the introduction and the discussion parts of the manuscript. I. Baek and S. Han wrote the manuscript with the help of S. Cheon and H.-W.Lee. All authors discussed the theoretical calculation and commented on the manuscript.

\noindent {\bf Acknowledgements}

We thank Cong Xiao for the fruitful advice on understanding and the calculation of the result and Hojun Lee, Daegeun Jo, and Byeonghyeon Choi for the fruitful advice for the first-principle calculation. I. Baek, S. Han, S. Cheon, and H.-W. Lee were supported by the Samsung Science and Technology Foundation (Grant No. BA-1501-51). Supercomputing resources, including technical support, were provided by the Supercomputing Center, Korea Institute of Science and Technology Information (Contract No. KSC-2022-CRE-0468).

\noindent {\bf Competing Interests}

The authors declare no competing interests.

\noindent {\bf Additional Information}
Correspondence and requests for materials should be addressed to H.-W.L.

\def\bibsection{\section*{References}}

\end{document}